\documentclass[pra, twocolumn,showpacs]{revtex4}  
\usepackage{graphicx}
\usepackage{amssymb}

\def\d#1{#1^\dagger}
 
\def\ket#1{|#1\rangle}

\begin{document}

\title{Multipartite continuous-variable entanglement from concurrent nonlinearities} 
\author{Olivier Pfister}
\email[Corresponding author: ]{opfister@virginia.edu}
\author{Sheng Feng}
\author{Gregory Jennings}
\author{Raphael Pooser}
\author{Daruo Xie}
\affiliation{Department of Physics, University of Virginia, 382 McCormick Road, Charlottesville, VA 22904-4714, USA}
\begin{abstract}
We show theoretically that concurrent interactions in a second-order nonlinear medium placed inside an optical resonator can generate multipartite entanglement between the resonator modes. We show that there is a mathematical connexion between this system and van Loock and Braunstein's proposal for entangling $N$ continuous quantum optical variables by interfering the outputs of $N$ degenerate optical parametric amplifiers (OPA) at a $N$-port beam splitter. Our configuration, however, requires only one nondegenerate OPA and no interferometer. In a preliminary experimental study, we observe the concurrence of the appropriate interactions in periodically poled $\rm RbTiOAsO_4$.
\end{abstract}
\maketitle

A crucial direction of effort in quantum information is the entanglement of (many) more than two systems. The current record number of entangled q-bits is four \cite{djw}. Continuous variables (CV) are a fascinating alternative to discrete ones and lend themselves well to quantum optical implementation \cite{book}. Multipartite entanglement of continuous variables was proposed by van Loock and Braunstein \cite{vLB,vLB2} and experimentally demonstated in two different regimes \cite{jing,aoki}. A  CV multipartite entangled state is an inseparable multimode squeezed state that tends towards a Greenberger-Horne-Zeilinger (GHZ) state in the limit of infinite squeezing. In van Loock and Braunstein's method, such states are generated from $N$ squeezed modes of the field emitted by optical parametric oscillators (OPO's) below threshold [i.e.\ optical parametric amplifiers (OPA's)] and combined by appropriately balanced beam splitters. The quantum interference at the beam splitters requires interferometric stabilization of the optical paths and indistinguishability, i.e.\ frequency and polarization degeneracy, of all $N$ modes. In this paper we show that multipartite entanglement is obtainable by use of a single OPA and no beam splitters. The OPA nonlinear medium must simultaneously phase-match several second-order nonlinearities that create two-mode squeezing between $N$ cavity field modes. The advantage of this scheme for experimental purposes is that it can be made very compact in a single periodically poled ferroelectric nonlinear crystal, with no interferometer to lock. Moreover, there is no degeneracy constraint on the frequencies. In the next section of this paper, we expose the theoretical arguments that prove our assertion. We then detail how a concurrence of three nonlinearities, suitable for entangling four modes, can be created in periodically poled ferroelectrics and we present preliminary observations of such a concurrence in periodically poled $\rm RbTiOAsO_4$ (PPRTA).

It is well known that a two-mode squeezer such as a nondegenerate OPA can generate an EPR state out of vacuum or coherent input \cite{reid,ou}.
The Hamiltonian in the interaction picture is
\begin{equation}\label{h1}
H_1 = i\hbar \beta\chi ( \d a_1 \d a_2 - a_1a_2),
\end{equation}
where $\chi$ is the nonlinear coupling coefficient and $\beta$ the real, assumed undepleted, coherent pump field amplitude. Solving the Heisenberg equations for the fields gives the squeezed joint quadratures
\begin{eqnarray}
P_1(t) + P_2(t) & = & (P_1 + P_2) e^{-\beta\chi t} \label{epr1}\\
X_1(t) - X_2(t)  & = & (X_1 - X_2) e^{-\beta\chi t} \label{epr2}
\end{eqnarray}
where $X=(a+\d a)/\sqrt 2$, $P=i(\d a-a)/\sqrt 2$, and $a(t=0)=a$. These EPR operators commute and admit maximally entangled common eigenstates such as
\begin{equation}
\int \ket x_1\ket x_2\ dx = \int \ket p_1\ket{-p}_2\ dp = \sum_{n=0}^{\infty} \ket n_1\ket n_2
\end{equation}
in the limit of infinite squeezing $\beta\chi t\rightarrow\infty$ \cite{wm,vE}.
We now ask whether concurrent interactions involving three modes would yield a tripartite CV entangled state, e.g.\ $\int\ket{xxx}dx$. It is simple to check that the Hamiltonian
\begin{equation}
H_2 = i\hbar \beta\chi_a ( \d a_1 \d a_2 - a_1a_2) + i\hbar \beta\chi_b ( \d a_2 \d a_3 - a_2a_3),
\end{equation}
does not create such CV tripartite entanglement: the solutions of the Heisenberg system yield only two squeezed joint modes out of three, the third one being a constant of motion initially subject to vacuum fluctuations. More interesting is the symmetrized three-mode Hamiltonian (we now take the interaction strengths equal, for the sake of simplicity and symmetry)
\begin{equation}
H_3 = i\hbar\beta\chi ( \d a_1 \d a_2 +   \d a_2\d a_3 +  \d a_3\d a_1) + H.c.,\label{h3}
\end{equation}
whose system of  Heisenberg equations is $\dot A={\cal M}\d A$, where $A^T=(a_1,a_2,a_3)$. ${\cal M}$ has only zeroes on the diagonal, $\beta\chi$ everywhere else, and eigenvalues $(2\beta\chi, -\beta\chi, -\beta\chi)$. The eigenmodes are joint operators 
\begin{eqnarray}\label{ghz}
P_1(t) + P_2(t) +P_3(t) &=&  (P_1 + P_2 + P_3) e^{-2\beta\chi t} \\
X_1(t) - X_2(t)  &=&  (X_1 - X_2) e^{-\beta\chi t} \\
X_1(t) - X_3(t) &=&  (X_1 - X_3) e^{-\beta\chi t},
\end{eqnarray}
whose common eigenstate is a multipartite entangled state that tends towards the GHZ state $\int |xxx\rangle dx$ when $\beta\chi t\rightarrow\infty$.  It is straightforward to generalize to
\begin{equation}
H_N = i\hbar\beta\chi \sum_{i=1}^N\sum_{j>i}^N \d a_i \d a_j + H.c.,\label{hn}
\end{equation}
which gives the set of multipartite entangled modes
\begin{eqnarray}
\sum_{i=1}^N P_i(t) &=&  e^{-(N-1)\beta\chi t} \sum_{i=1}^N P_i\label{ghz1}\\
X_i(t)-X_j(t)  &=&  e^{-\beta\chi t} (X_i-X_j),\ \forall i\neq j.\label{ghz2}
\end{eqnarray}
(Note that the phase sum squeezing is higher than for the other GHZ modes.) It would therefore seem that we have found a procedure for entangling an arbitrary number of modes: one just needs to have every one of them interacting equally strongly with all the others. Before we turn to the experimental feasibility of this scheme, we examine its relation to that of van Loock and Braunstein.

A connexion is already known for bipartite entanglement. Indeed, $H_1$ [Eq.\ (\ref{h1})] can also be viewed as the result of the transformation of the Hamiltonian of two single-mode squeezers, $H'_1= i \frac{\hbar}2 \beta\chi [-(a_1^{2\dagger} - a_1^2)+ ( a_2^{2\dagger} - a_2^2)]$, by a lossless balanced beam splitter: $H''_1=U_{BS}H'_1\d U_{BS}=H_1$ \cite{ll}. Both possibilities have been used experimentally: $H_1$ in the first CV EPR experiment \cite{ou} and subsequent ones \cite{pengEPR,schori} and $H''_1$ in quantum teleportation experiments \cite{furu,lam}.

We now show the same kind of simple relationship exists for CV multipartite entangled beams. The original proposal of CV multipartite entanglement uses three single-mode squeezed beams, produced by $H'_3 = i \frac{\hbar}2 \beta\chi (-a_1^{2\dagger} + a_2^{2\dagger} + a_3^{2\dagger})+ H.c.$ and  mixed by a ``tritter", i.e.\ the combination of a 2:1 and a 1:1 beamsplitter \cite{vLB,vLB2}. The transformation of $H'_3$ by the tritter is
\begin{equation}
H''_3 =  \frac 1 3\left[ i \frac{\hbar\beta\chi}{2}  (b_1^{2\dagger} +b_2^{2\dagger} +b_3^{2\dagger})+H.c. - 2 H_3 \right].
\end{equation}
i.e.\ a combination of our $H_3$ [Eq.\ (\ref{h3})] and symmetrized single-mode squeezers. The two-mode dependence of $H''_3$ and $H_3$ is therefore identical.

The corresponding $N$-mode entangling Hamiltonian is
\begin{equation}\label{HvLB}
H''_N =   i\hbar\beta\chi \left[\frac{N-2}{2N} \sum_{i=1}^N b_i^{2\dagger}\right] +H.c.  - \frac 2 NH_N,
\end{equation}
where $H_N$ is our entangling Hamiltonian [Eq.\ (\ref{hn})]. The system of Heisenberg equations for $H''_N$ is $\dot B={\cal M}\d B$, where $B^T=(b_1,\dots,b_N)$. All of ${\cal M}$'s diagonal elements equal $(N-2)\beta\chi/N$ and all off-diagonal elements equal $-2\beta\chi/N$. The eigenvalues are $(-\beta\chi,\beta\chi,\dots,\beta\chi)$ and eigenvectors of the form of Eqs.\ (\ref{ghz1},\ref{ghz2}), with $X$ and $P$ swapped and equal squeezing rates $\exp(-\beta\chi t)$. Our Hamiltonian $H_N$ (or $H_3$) leads to the same matrix, but with zero diagonal, and asymmetric squeezing rates [Eqs.\ (\ref{ghz1},\ref{ghz2})]. The advantage is that no control is required over the phases of the input squeezed states, i.e.\ over the signs of the different terms in $H'_{1,3,N}$. This greatly simplifies the experimental setup, as one passes from the interference of $N$ OPA's to the output of a single OPA.

The experimental principle is to use the simultaneous nonlinear interaction of different eigenmodes of an optical resonator. The presence of the resonator here simplifies the situation by selecting a discrete comb of resonant modes out of the quantum field continuum. Moreover, because the assembly of the optical resonator with the nonlinear medium constitutes a parametric oscillator, it provides us with the option of operating either in the spontaneous emission regime (below oscillation threshold) or in the stimulated emission regime (above oscillation threshold). We plan to investigate the latter case in subsequent work but we restrict the scope of this paper to vacuum-seeded optical parametric amplification.

In general, a nonlinear second-order medium pumped at frequency $\omega_p$ will optimally couple pairs of modes $\omega_{1,2}$ such that $\omega_1+\omega_2=\omega_p$ (from the phase-matching condition). The phase-matching bandwidth can be broad enough (e.g.\ 100 GHz) for several pairs of modes, separated by a free spectral range (FSR), say, $\Delta\leq 1$ GHz, to have approximately the same coupling strength. It is also well known that the resonance condition depends on the frequency, via dispersion, but these effects are small enough to be negligible over a few free spectral ranges. Still, a {\em singly-pumped} OPA cannot realize multipartite entanglement. The two possible cases are:

{\em (i) $\omega_p$ coincides with twice a cavity resonance frequency $2\omega_0$.} Then the Hamiltonian is
\begin{equation}\label{Hsdeg}
i\hbar\beta\chi\left(\frac 1 2 a_0^{2\dagger} + \d a_1\d a_{-1} + \d a_2\d a_{-2}+\cdots\right) + H.c.
\end{equation}
where $a_0$ has frequency $\omega_0$ and $a_{\pm k}$, $\omega_{\pm k}=\omega_0+\pm k\Delta$.

{\em (ii) $\omega_0'$ coincides with the sum of two consecutive cavity resonance frequencies,} e.g.\ $\omega_0'=\omega_0+\omega_1$. The Hamiltonian is
\begin{equation}\label{Hsndeg}
i\hbar\beta\chi\left( \d a_0\d a_1 + \d a_{-1}\d a_{2}+\d a_{-2}\d a_{3}+\ldots\right) + H.c.
\end{equation}

In neither case is a multipartite entangling Hamiltonian realized because we never have a given set of more than two modes all connected together by the interaction. This problem can be easily solved by having several pump beams at different frequencies, more precisely half a FSR apart. This amounts to adding both Hamiltonians (\ref{Hsdeg}) and (\ref{Hsndeg}). However, an additional difficulty stems from the fact that the degenerate interaction in (\ref{Hsdeg}) has the same sign as the other nondegenerate interactions, contrary to what is required by Eq.\ (\ref{HvLB}). In fact, it is impossible to have a different interaction phase for two different downconversion terms that share the same pump such as in
\begin{equation}
i\hbar\beta\chi\left(-\frac 1 2 a_0^{2\dagger} + \d a_1\d a_{-1} + \d a_2\d a_{-2}+\ldots\right) + H.c.,
\end{equation}
to be compared with Eq.\ (\ref{Hsdeg}). This makes the {\em exact} realization of van Loock and Braunstein's Hamiltonian $H''_N$ impossible with concurrent interactions, as it requires opposite signs for the degenerate and nondegenerate interactions [Eq.\ (\ref{HvLB})]. Thus, one must go back to the other side of the multiport splitter, where the individually tunable optical paths provide the necessary degrees of freedom. However, concurrent interactions can realize our alternate, nondegenerate Hamiltonian $H_N$  [Eq.\ (\ref{hn})]. 

\begin{figure}[htb]
\begin{center}
\begin{tabular}{c}
\includegraphics[width= 2.9in]{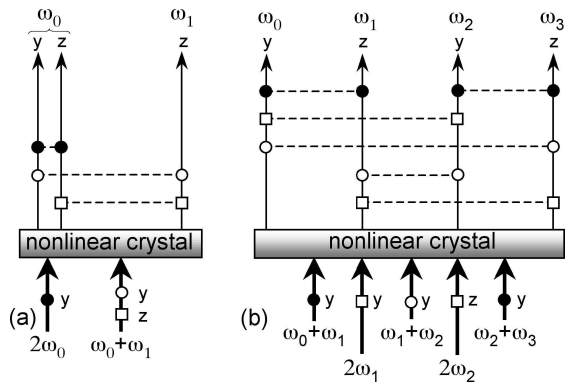}
\end{tabular}
\end{center}
\caption{concurrent entangling interactions (dashed lines) for (a): three and (b): four OPA modes (top arrows). Bottom arrows represent the pump fields.}
\label{ent}
\end{figure}

In practice, the implementation of $H_N$ requires obtaining all possible nondegenerate coupling terms without any degenerate one, for a given set of modes. We now show how this can be achieved for sets of three or four modes. Several equivalent possibilities exist. Without tediously enumerating them all, we focus, without loss of generality,  on the simplest ones that are experimentally realizable. These are the only ways we have found to fulfill the two requirements stated above (which does not, obviously, constitute a proof of unicity). We assume propagation along the principal axis $x$ of a nonlinear crystal and use the polarization ($y$, $z$) and frequency degrees of freedom to label the modes. Graphical representations of the interactions are given on Fig.\ \ref{ent}.
Three-mode entanglement is realized by simultaneously phase-matching $zzz$ and $yzy$ parametric downconversion [Fig.\ \ref{ent}(a)]:
\begin{eqnarray}
H_3 = &i\hbar&\left[\beta_y(2\omega_0)\chi_{yzy}  \d a_y(\omega_0) \d a_z(\omega_0)\right.\nonumber\\
&+& \beta_y(\omega_0+\omega_1)\chi_{yzy}  \d a_y(\omega_0) \d a_z(\omega_1) \nonumber\\
&+&\left. \beta_z(\omega_0+\omega_1)\chi_{zzz}  \d a_z(\omega_0) \d a_z(\omega_1)\right] + H.c.\label{h3exp}
\end{eqnarray}
No other interaction occurs inside the set of modes $\{a_y(\omega_0), a_z(\omega_0),a_z(\omega_1)\}$ [in particular, $yyy$ is not phase-matched and there is no $\beta_z(2\omega_0)$]. The pump amplitudes can be adjusted to make up for residual differences in the nonlinear coefficients $\chi_{ijk}$ so that the interaction strengths are all equal. Under these conditions, this nondegenerate OPA should amplify vacuum inputs into tripartite entangled modes of the form (\ref{ghz}). The four-field entanglement Hamiltonian $H_4$ necessitates additional phase-matching of the $yyy$ interaction, which can be done by using a $zzz$ crystal rotated by $\rm 90^o$. Neglecting the FSR difference between the two polarizations (which, for practical purposes, can be made arbitrarily small using two crystals) we can design an entangling interaction between four equally spaced cavity modes [Fig.\ \ref{ent}(b)]:
\begin{eqnarray}
&H_4& = i\hbar\left\{\beta_y(\omega_0+\omega_1)\chi_{yzy}  \d a_y(\omega_0) \d a_z(\omega_1)\right. \nonumber\\
&+& \beta_y(2\omega_1)\chi_{yyy}  \d a_y(\omega_0) \d a_y(\omega_2)\nonumber\\
&+& \beta_y(\omega_1+\omega_2)\chi_{yzy} \left[ \d a_z(\omega_1) \d a_y(\omega_2)+ \d a_y(\omega_0) \d a_z(\omega_3)\right] \nonumber\\
&+& \beta_z(2\omega_2)\chi_{zzz}  \d a_z(\omega_1) \d a_z(\omega_3) \nonumber\\
&+&\left. \beta_y(\omega_2+\omega_3)\chi_{yzy}  \d a_y(\omega_2) \d a_z(\omega_3)\right\} + H.c.
\end{eqnarray}
Again, these are all the possible pair couplings inside the set of modes $\{a_y(\omega_0), a_z(\omega_1),a_y(\omega_2),a_z(\omega_3)\}$ and these should thus all be entangled by $H_4$. Note that the absence of $yzz$ and $zyy$ prevents degenerate terms from appearing. The equidistant pump frequencies can easily be obtained by acousto-optic or electro-optic modulation. This scheme does not scale easily to $N>4$, unfortunately, because this requires phase-matching $zyy$/$yzz$ and it then becomes impossible to avoid degenerate interactions with the wrong sign, which, we have found, always dramatically reduce the number of entangled modes (detailed calculations will be published elsewhere). This is a consequence of the freezing of the optical phases in our scheme; however, this very limitation is precisely what makes possible the experimental simplicity of our approach, which could lead to extremely compact and low-loss CV multipartite entanglers.

As we have seen, realizing $H_{3,4}$ in a nonlinear optical material necessitates the concurrence of two different interactions. Using two different nonlinear crystals is a possibility but it is costly in terms of optical losses, especially inside an optical resonator. It is therefore desirable, in line with the rationale of this paper, to obtain all interactions in the same crystal. While such coincidences are extremely rare in birefringent phase-matching, they are very easy to engineer in quasi-phase-matched materials. Quasi-phase-matching (QPM) is as old as nonlinear optics \cite{qpm}. It relies on spatial modulation of the nonlinear coefficient, here by periodically poling a ferroelectric crystal, to make up for the phase mismatch of a particular nonlinear interaction \cite{fejer}. QPM allows practically any interaction to be phase-matched in the same material by simply changing the poling period. Poling the same crystal with two different periods therefore yields the desired coincidence. Even simpler designs are possible: one can find two interactions sharing the same period \cite{ol97}. Moreover, since the spatial modulation of the nonlinear coefficient is a square wave, one can also find coincidences between different poling harmonics. We have observed one instance of these in PPRTA. RTA is an isomer of KTP with lower residual absorption losses and equivalent nonlinear coefficients. Like KTP, RTA is non-hygroscopic, has a very high optical damage threshold, and presents neither photorefractive damage nor blue-induced infrared absorption. We consider the particular set of Nd-doped laser wavelengths of 532 nm, for the pump fields, and 1064 nm, for the parametrically amplified fields. The poling periods required to quasi-phase-match SHG of 1064 nm at room temperature are 43 $\mu$m for $yzy$ and 8.37 $\mu$m for $zzz$. The interaction is tunable by varying the refractive indices via the wavelength, incidence angle, or temperature. 
\begin{figure}[t]
\begin{center}
\begin{tabular}{c}
\includegraphics[width= 2.95in]{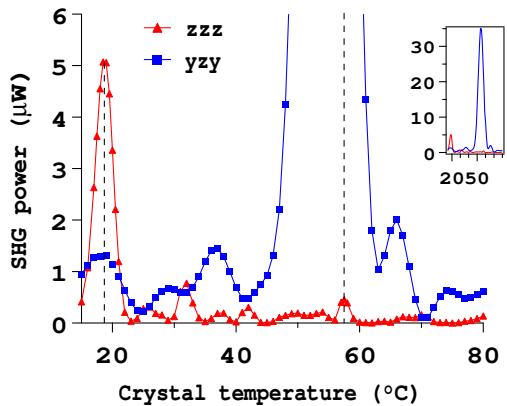}
\end{tabular}
\end{center}
\caption{Simultaneous quasi-phase-matching of $yzy$ and $zzz$ SHG in PPRTA. The SHG powers are measured versus crystal temperature for the zzz and yzy interactions. Concurrences involving secondary maxima are at $\rm 18.75^oC$ and $\rm 57.5^oC$ (dashed lines). Inset: full-scale data plot.}
\label{casc}
\end{figure} 
The goal is to obtain simultaneous QPM of both $yzy$ and $zzz$ interactions at the same temperature. The 8.37 $\mu$m period is the fifth harmonic of 41.85 $\mu$m, close to 43 $\mu$m. We must therefore find an intermediate poling period for which both interactions occur at the same temperature. The temperature dependance of RTA indices being not perfectly known yet, we have tried a 41.95 $\mu$m crystal. The results of SHG measurements are plotted in Fig.\ \ref{casc}. The input beam at 1064 nm and the output beam at 532 nm are respectively polarized and analyzed with polarizers. Even though the two main QPM peaks are still separated by about $\rm 40^o$C, we already have coincidences of each main peak of one interaction with a secondary one of the other interaction. The ratio of the two maxima gives $(d^{\mathit{eff}}_{yzy}/d^{\mathit{eff}}_{zzz})^2 = 7$, for a theoretical value of $(5d_{24}/d_{33})^2\simeq 1.8$. We attribute the discrepancy to high frequency irregularities in the crystal's poling, which would affect the fifth harmonic of the period more than its fundamental. We are currently refining measurements of temperature tuning for RTA in order to improve the period design.

In conclusion, we have demonstrated that concurrent nonlinearities in a single OPA can entangle four modes, possibly more. The current technology makes such an OPA quite feasible and we are now preparing an experimental realization. The compactness and simplicity of such a source of entangled beams should make it very attractive for CV quantum information implementations, such as teleportation networks, controlled dense coding, and quantum error correction via telecloning. 

We thank Richard Stolzenberger and Tito Ayguaviv\`es, both formerly of Coherent Crystal Associates, Inc., for their help with PPRTA. This work is supported by NSF grants
PHY-0245032 and EIA-0323623. 

{\em Note added in proof}: we have become aware of related work for tripartite photon-number entanglement \cite{mp}.

 \end{document}